\documentclass[aps, twocolumn, showpacs, amsmath, amssymb]{revtex4}
\usepackage{graphicx}
\usepackage{amsmath}
\usepackage{amsfonts}
\newcommand{\Sil}{SiO$_2$}
\newcommand{\Tan}{Ta$_2$O$_5$}
\newcommand{\TanSil}{Ta$_2$O$_5$/SiO$_2$}
\newcommand{\TanAl}{Ta$_2$O$_5$/Al$_2$O$_3$}
\begin{document}
\title{Mechanical Loss in Tantala/Silica Dielectric Mirror Coatings}
\author{Steven D. Penn$^{1,}$\footnote{Present address: Department of Physics, Hobart and William Smith Colleges, Geneva, New York, 14456, USA.}, Peter H. Sneddon$^{2}$, Helena Armandula$^{3}$, Joseph C. Betzwieser$^{4}$, Gianpietro Cagnoli$^{2}$, Jordan Camp$^{3}$\footnote{Present address: Laboratory for High Energy Astrophysics, NASA/Goddard Space Flight Center, Greenbelt, Maryland, 20771 USA.}, D. R. M. Crooks$^{2}$, Martin M. Fejer$^{5}$, Andri M. Gretarsson$^{1}$\footnote{Present address: LIGO Livingston Observatory, Livingston, Louisiana, 70754 USA.},  Gregory M. Harry$^{4}$, Jim Hough$^{2}$, Scott E. Kittelberger$^{1}$, Michael J. Mortonson$^{4}$, Roger Route$^{5}$, Sheila Rowan$^{5}$, Christophoros C. Vassiliou$^{4}$ \vspace{3ex}}
\affiliation{$^{1}$Department of Physics, Syracuse University, Syracuse, New York 13244-1130,  USA.\\
$^{2}$Department of Physics and Astronomy, University of Glasgow, Glasgow G12 8QQ, Scotland, United Kingdom.\\
$^{3}$LIGO Laboratory, California Institute of Technology, Pasadena, California 91025, USA.\\
$^{4}$LIGO Laboratory, Massachusetts Institute of Technology, Cambridge, Massachusetts 02139, USA.\\
$^{5}$Edward L. Ginzton Laboratory, Stanford University, Stanford, California 94305-4085, USA\vspace{3ex}}
\date{\today}
\begin{abstract}\label{Abstract}
\noindent Current interferometric gravitational wave detectors use test masses
with mirror coatings formed from multiple layers of dielectric materials, most
commonly alternating layers of \Sil\ (silica) and \Tan\ (tantala). However,
mechanical loss in the \TanSil\ coatings may limit the design sensitivity for
advanced detectors. We have investigated sources of mechanical loss in the
\TanSil\ coatings, including loss associated with the coating-substrate
interface, with the coating-layer interfaces, and with the bulk material. Our
results indicate that the loss is associated with the bulk coating materials
and that the loss of \Tan\ is substantially larger than that of \Sil.
\end{abstract}
\maketitle
\section{Introduction}\label{Introduction}
The sensitivity of designs for advanced interferometric gravitational wave
detectors, such as Advanced LIGO (Laser Interferometer Gravitational Wave
Observatory), is limited in the frequency range from 10's to 100's of Hz by
thermal noise from the main test masses and their suspensions~\cite{AdvDesign}.
 These test masses, which under current design plans will be formed from either
fused silica or sapphire, will each be supported using a fused silica
suspension, and will have multi-layer, dielectric mirror
coatings~\cite{whitepaper}. In the current generation of gravitational wave
detectors, the test mass mirror coatings are formed by ion-sputtering
alternating layers of silicon dioxide (\Sil) and tantalum pentoxide
(\Tan).
This type of coating was chosen because it can be
made highly reflective in a narrow band around 1.064 $\mu$m, the laser
wavelength chosen for LIGO, while having very low absorption and scatter
losses~\cite{CurrentCoatings}.

For the past several years, many research groups in the gravitational wave
field have been concerned that dielectric coatings could be an important source
of thermal noise. This concern was strengthened by Levin's
calculation~\cite{Levin} which indicated that mechanical losses in the mirror
surface of a test mass could be much greater than had been generally
appreciated. Subsequently, our investigations of mirror coatings showed that
multi-layers of \TanSil\ and \TanAl\ when applied to fused silica substrates,
add significant levels of mechanical loss~\cite{Crooks,Harry02}.  Using these
results and models developed by Nakagawa~\cite{Nak02}, and
Gretarsson~\cite{Harry02}, we calculated that mechanical loss in the coatings
would result in a level of thermal noise which would degrade the design
sensitivity of the planned advanced detectors by a significant amount.

In the present study, we have performed a series of experiments to investigate
the source of the loss in multi-layer tantala/silica coatings. We hypothesized
that the loss would arise primarily either from the coating-substrate
interface, from the coating-layer interfaces, from the bulk materials used in
the coatings, or possibly from some combination of these sources. We have
measured the loss of a series of coatings in which the number and thickness of
the coating layers was chosen to test each of these dependencies. 

\section{Design of coating loss study}

\begin{figure}[htbp]
\includegraphics[width=8cm]{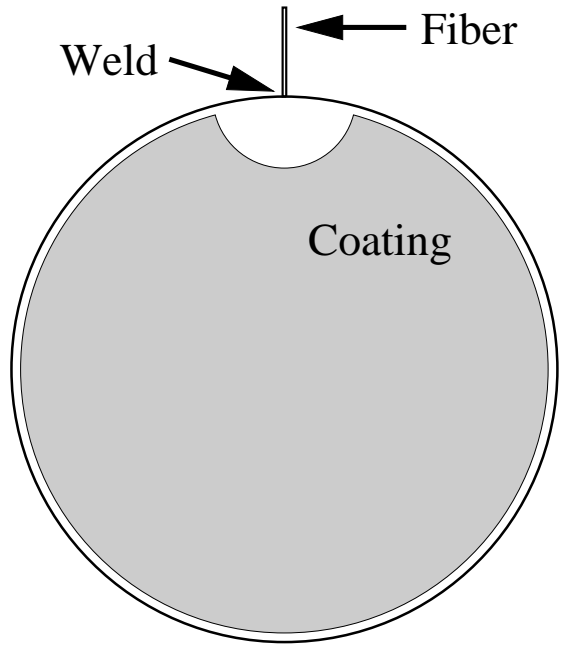}
\caption{Coating on Thin sample with semicircular mask around weld.}
\label{Fig_ThinSample}
\end{figure}

\subsection{Experimental technique}

As in our previous experiments~\cite{Crooks,Harry02}, we determine the level of
mechanical loss associated with a given coating by applying the coating to
fused silica substrates and measuring the mechanical losses of a subset of
resonant modes of the coated samples. 

The mechanical loss, $\phi$, at a resonant frequency $f_0$ of a sample, is
related to the quality factor, $Q$ of the resonance by $\phi(f_0)=1/Q$.
Assuming all other losses to be negligible, the total loss in a coated sample,
$\phi_\mathrm{coated}(f_0)$, is equal to the sum of the intrinsic loss of the
substrate plus any loss associated with the coating~\cite{Crooks,Harry02},
\begin{equation}\label{Eqn_PhiCoated}
\phi_\mathrm{coated}(f_0) \approx \phi_\mathrm{substrate}(f_0) + \frac{E_\mathrm{coating}}{E_\mathrm{substrate}} \phi_\mathrm{coating}(f_0)
\end{equation}
where ${E_\mathrm{coating}}/{E_\mathrm{substrate}}$ is the ratio of energy
stored in the coating to energy stored in the substrate.

In this study, the substrates were 7.6 cm diameter fused silica disks of two
different thicknesses, 2.5 cm and 0.25 cm. The thinner disks had the advantage
that the effect from the coating was more pronounced; the coating loss had a
greater contribution to the total loss because the energy ratio,
${E_\mathrm{coating}}/{E_\mathrm{substrate}}$, was larger for the thin samples.
However suspending these thin samples required the direct welding of a low loss
silica suspension to the edge of the fused silica substrates, see
\cite{Harry02}. To minimize any potential damage to the coatings by direct heat
from welding, the coating was masked in a 1 cm radius around the weld, see
figure~\ref{Fig_ThinSample}.  In addition, we used miniature welding torches
that allowed the weld area to be sub-millimeter in scale. During welding, the
glass $\approx 1$ cm from the weld remained only warm to the touch, which
suggests that there was little to no impact on the coating. Although
measurements of the loss factors of the thicker samples were less sensitive to
the effects of coating losses, they had the advantage that they could be
suspended using thin silk thread, see \cite{Crooks}, without any risk of
physically altering the substrate or the coating. 

The use of both techniques allowed a more thorough investigation of mechanical
losses in the coating over a wider range of frequencies~\cite{SneddonRowan}.

The substrates were made of Corning 7980 grade 0A fused silica~\cite{Corning}. Their faces were polished to sub-angstrom micro-roughness by Wave Precision
Inc.~\cite{WavePrecision}, to emulate the required surface properties of actual
test masses. Except where otherwise noted, all the coatings were applied by SMA/Virgo~\cite{SMA} at {\em l'Institut de Physique Nucleaire} in Lyon France.

To measure the quality factor of a given mode of a sample, we suspend the
sample in vacuum and excite it to resonance using an electrostatic drive. We
then remove the excitation signal and record the sample's motion as it freely
rings down. The characteristic decay time, $\tau$, is the time required for the
amplitude of motion to decrease by a factor $1/e$. The quality factor is then
given by $Q=\pi f_0 \tau$. The amplitude of the resonant motion was sensed
interferometrically for the thick ($2.5$ cm) samples, and by polarimetry for
the thin ($0.25$ cm) samples. At the vacuum pressure for the experiment, ($P \leq 10^{-6}$ torr), air damping was negligible. Detailed descriptions of these ringdown
techniques are given in ~\cite{Crooks} and ~\cite{Harry02} respectively.

We performed finite element analysis, with the program {\tt
ALGOR}~\cite{ALGOR},  to model the displacement of each mode of the samples.
The relevant energy ratios for the coated samples were then calculated from the
displacements~\cite{Crooks}. If the intrinsic mechanical loss of the substrate,
$\phi_\mathrm{substrate}(f_0)$, is known (or is insignificant compared to the
effective loss from the applied coating), the loss associated with the coating
may then be calculated using eqn.~\ref{Eqn_PhiCoated}.

\begin {table}[htbp]
\begin{tabular}{|c||c|c|c|c|}   \hline
Sample & Total  & \multicolumn{2}{c|}{Optical Thickness} & \\
Type   & Layers & ~~~\Sil\ ~~~ & \Tan\ & Comments \\ \hline\hline
A & 0  & 0 & 0 & Annealed only \\ \hline
B & 2  & $\lambda/4$ & $\lambda/4$ & Coated \& annealed \\ \hline
C & 30 & $\lambda/4$ & $\lambda/4$ & Coated \& annealed  \\ \hline
D & 60 & $\lambda/8$ & $\lambda/8$ & Coated \& annealed  \\ \hline
E & 30 & $\lambda/8$ & ${3\lambda}/8$ & Coated \& annealed  \\ \hline
F & 30 & ${3\lambda}/8$ & $\lambda/8$ & Coated \& annealed  \\ \hline
\end{tabular}
\caption{Set of samples used to probe the source of the mechanical loss in the \TanSil\ coating.}
\label{Tbl_SampleTypes}
\end{table}

\subsection{Sequence of coatings studied}
We investigated three primary sources of mechanical loss in the coatings, which we postulated to be:
\begin{itemize}
   \item loss in the coating-substrate interface,
   \item loss in the interfaces between the multiple coating layers, and  
   \item loss in the bulk coating materials.  
\end{itemize}

To investigate these hypotheses, we designed a set of coatings in which we
varied the number and thickness of the coating layers to test for the three
dependencies.  Table~\ref{Tbl_SampleTypes} lists the coatings and the substrate
treatments investigated.  By comparing the measured mechanical loss for sample
types B through F, we can test our three hypotheses for the source of the
coating loss.

If the dominant source of coating mechanical loss is associated with the
coating-substrate interface, then the total loss measured for sample types BÐ-F
would be expected to be approximately equal.

On the other hand, the loss may have originated predominantly in the interfaces
between the multiple coating layers. In that case the coating loss of sample
type D, which has 60 layers, should be twice as large as the loss in sample
types C, E and F which have 30 layers.

Finally, the source of dissipation may be intrinsic to the bulk coating
material. Under that scenario, the coating loss for sample types C, E, and F
should vary with the abundances of the two coating materials.

Clearly there also exists the possibility that the total mechanical loss of the
coating is the sum of contributions from a combination of the above mechanisms.

As part of the coating process, samples are heated to a few hundred degrees to
reduce residual stresses in the coating. Previous experiments~\cite{Numata,
Lunin, Penn} have shown that heating uncoated fused silica samples can result
in a significant decrease in loss. Coating Type A was not coated but was
cleaned and heated to the same elevated temperature as the coated samples to
ensure any change in substrate loss was taken into account in our analysis.  We
will refer to this process as ``annealing'' even though the samples were not
raised to the standard annealing temperatures for fused silica.

\begin{figure}[htbp]
\includegraphics[width=8cm]{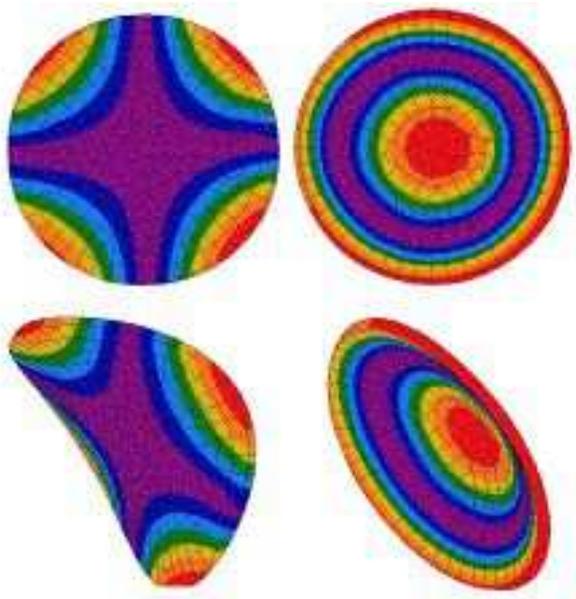}
\caption{Modes of the thin sample. The first mode (Butterfly) on the left and the second mode (Drumhead) on the right. The first row is face view and the second row is side view.}
\label{Fig_ThinModes}
\end{figure}

\begin {table*}[tbp]
\begin{tabular}{|c|c||c|c|c||c|c|c||c|c|c||c|c|}   \hline
\multicolumn{2}{|c||}{Sample}   & \multicolumn{3}{c||}{Butterfly $\times$ mode} & \multicolumn{3}{c||}{Butterfly $+$ mode} & \multicolumn{3}{c||}
{Drumhead mode} &  \multicolumn{2}{c|}{Coating Thickness} \\  
Type & Number &$Q_\mathrm{uncoated}$&$Q_\mathrm{coated}$&
$\phi_\mathrm{coating}$&$Q_\mathrm{uncoated}$&$Q_\mathrm{coated}$&$\phi_\mathrm{coating}$&$Q_\mathrm{uncoated}$&
$Q_\mathrm{coated}$&$\phi_\mathrm{coating}$&~~~$s_\mathrm{silica}$ ($\mu$m)~~~&$s_\mathrm{tantala}$ ($\mu$m) \\ \hline\hline
A & 1 & 14.7 &      & N/A  & 11.7 & 43.6 & N/A &  &      &     & \multicolumn{2}{c|}{Uncoated, Annealed at $600^\circ$ C} \\ \hline
A & 2 & 10.6 & 42.4 & N/A  & 13.9 & 54.0 & N/A &  &      &     & \multicolumn{2}{c|}{Uncoated, Annealed at $900^\circ$ C} \\ \hline
B & 1 & 24.2 & 9.0  & 2.4  & 25.2 & 8.0  & 2.7 &  &      &     & 0.183 & 0.131  \\ \hline
B & 2 & 6.32 & 5.4  & 4.0  & 7.53 & 6.5  & 3.3 &  & 6.4  & 3.2 & 0.183 & 0.131  \\ \hline
C & 1 & 18.4 & 0.53 & 2.7  & 20.8 & 0.55 & 2.6 &  &      &     & 2.75  & 1.97   \\ \hline
C & 2 & 23.1 & 0.53 & 2.7  & 18.6 & 0.54 & 2.6 &  & 0.43 & 3.1 & 2.75  & 1.97   \\ \hline
D & 1 & 17.3 & 0.49 & 2.9  & 20.4 & 0.55 & 2.6 &  & 0.44 & 3.1 & 2.75  & 1.97   \\ \hline
D & 2 &      & 0.54 & 2.6  & 43.6 & 0.51 & 2.8 &  &      &     & 2.75  & 1.97   \\ \hline
E & 1 & 42.4 & 0.40 & 3.9  & 54.0 & 0.40 & 3.9 &  & 0.29 & 5.1 & 1.38  & 2.95   \\ \hline
E & 2 & 22.2 & 0.40 & 3.9  & 20.3 & 0.41 & 3.8 &  &      &     & 1.38  & 2.95   \\ \hline
F & 1 &      & 0.75 & 1.8  &      & 0.72 & 1.8 &  &      &     & 4.13  & 0.983  \\ \hline
F & 2 &      & 1.13 & 1.2  &      & 0.82 & 1.6 &  & 0.63 & 2.0 & 4.13  & 0.983  \\ \hline
\end{tabular}
\caption{Results for thin sample measurements.  All $Q$'s are $\times 10^6$. All $\phi$'s are $\times 10^{-4}$.}
\label{Tbl_ThinResults}
\end{table*}

\section{Results}
\subsection{Thin Sample Results} \label{Sect_ThinResults}
We present below the results of our measurements on the thin samples.  We
acquired at least five measurements for each mode of a given sample
configuration. As is common in $Q$  experiments, we quote the result of the
highest measurement. The reason for this selection process is that systematic
errors, which in many experiments can equally increase or decrease one's
measurement, will, with rare exception in our measurements, uniformly degrade
the $Q$.  On the other hand, the statistical errors contribute a uniform,
Gaussian background that introduces an error to the fit of $Q$.  Data that does
not exhibit uniform background noise is discarded.  These statistical errors
are normally much smaller that the systematic fluctuations.  Therefore we quote
the largest value for our $Q$ measurements with its associated statistical
error.

The suspension system, excitation system, and data analysis routines have all
been tested to produce no limiting effect on measurements where the $Q$ exceeds
80 million.  During these measurements, the dimensions of the suspension system
and the position of the excitation system were altered with no observed change
in the $Q$ of the sample.  Finally these measurements were performed at MIT and
at Syracuse University with excellent agreement between the two facilities.

The results are listed by mode in Table~\ref{Tbl_ThinResults}. The mode shapes
are shown in Figure~\ref{Fig_ThinModes}. The Butterfly-$\times$ and
Butterfly-$+$ modes are the degenerate first mode with a typical resonant
frequency of $f_{1} = 2.7$ kHz.  The additional notation of ``$\times$'' and
``$+$'' refer to the orientation of the nodal lines.  The second mode, known as
the drumhead mode, has a typical resonant frequency of $f_2 = 4.1$ kHz.

One immediate observation from the data is that for the coated samples, with
the exception of the 2-layer coating, $Q_\mathrm{coated} <<
Q_\mathrm{uncoated}$.  Typical values for $Q_\mathrm{uncoated}$ indicate that
the loss in the substrate is about 2--3\% of $Q_\mathrm{coated}$ and should not
significantly affect the results for the coating loss. Moreover, we see that
the annealing performed at the end of the coating run dramatically increases
the $Q$ of an uncoated substrate.  If we assume that the substrate in a coated
sample also experiences the same increase in $Q$, then the loss in the
substrate is reduced to about 1\% of the loss in the coated sample.  Thus, for
the thin samples, we assume the loss measured in a coated sample is approximately
the loss associated with the coating.
\begin{equation}
\phi_\mathrm{coated} \approx
\frac{E_\mathrm{coating}}{E_\mathrm{substrate}} \phi_\mathrm{coating}
\end{equation}

To calculate the coating losses we require the fraction of the mode energy
stored in the coating.  This value is calculated from the displacement values
obtained using finite element analysis package, {\tt ALGOR}~\cite{ALGOR}.
This analysis yielded an energy ratio per coating thickness of 
\begin{eqnarray*}
\frac{{\mathrm{d}E}/{\mathrm{d}s}}{E} = 1488 \mathrm{~m}^{-1} \mathrm{~~~~~~Butterfly~mode} \\
\frac{{\mathrm{d}E}/{\mathrm{d}s}}{E} = 1575 \mathrm{~m}^{-1} \mathrm{~~~~Drumhead~mode} \\
\end{eqnarray*}

Then, given the coating thickness, $s$, we can rewrite the equation for
$\phi_\mathrm{coating}$ as
\begin{equation}\label{Eqn_PhiThin}
\phi_\mathrm{coated}(f_0) = s \frac{{\mathrm{d}E}/{\mathrm{d}s}}{E} \phi_\mathrm{coating}(f_0)
\end{equation}
The coatings are made such that the first layer applied to the substrate is a
tantala layer.  
We convert a coating's optical thickness to its
physical thickness using the refractive indices 
$n_\mathrm{silica} = 1.45$ and $n_\mathrm{tantala} = 2.03$.

\begin{figure}[htbp]
\includegraphics[width=8cm]{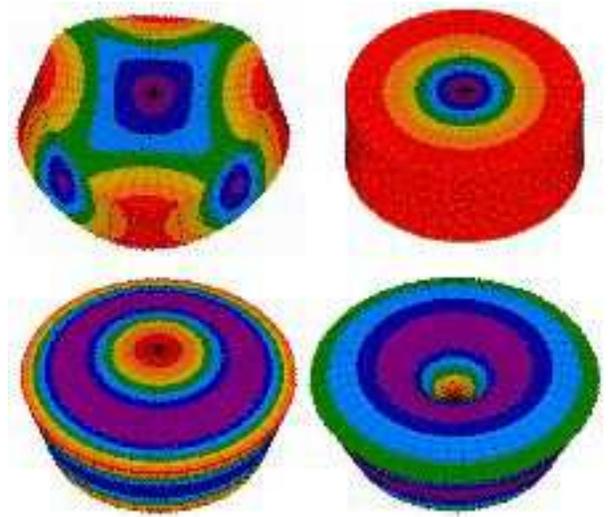}
\caption{The shapes of the four modes of the thick sample:  Clover-4 or Butterfly (C4) in upper left; Fundamental Radial (F) in upper right; Asymmetric Drumhead (A) in lower left, and $2^{\mathrm{nd}}$ Asymmetric (2A) in lower right.}
\label{Fig_ThickModes}
\end{figure}
 
\begin{figure*}
\includegraphics[width=18cm]{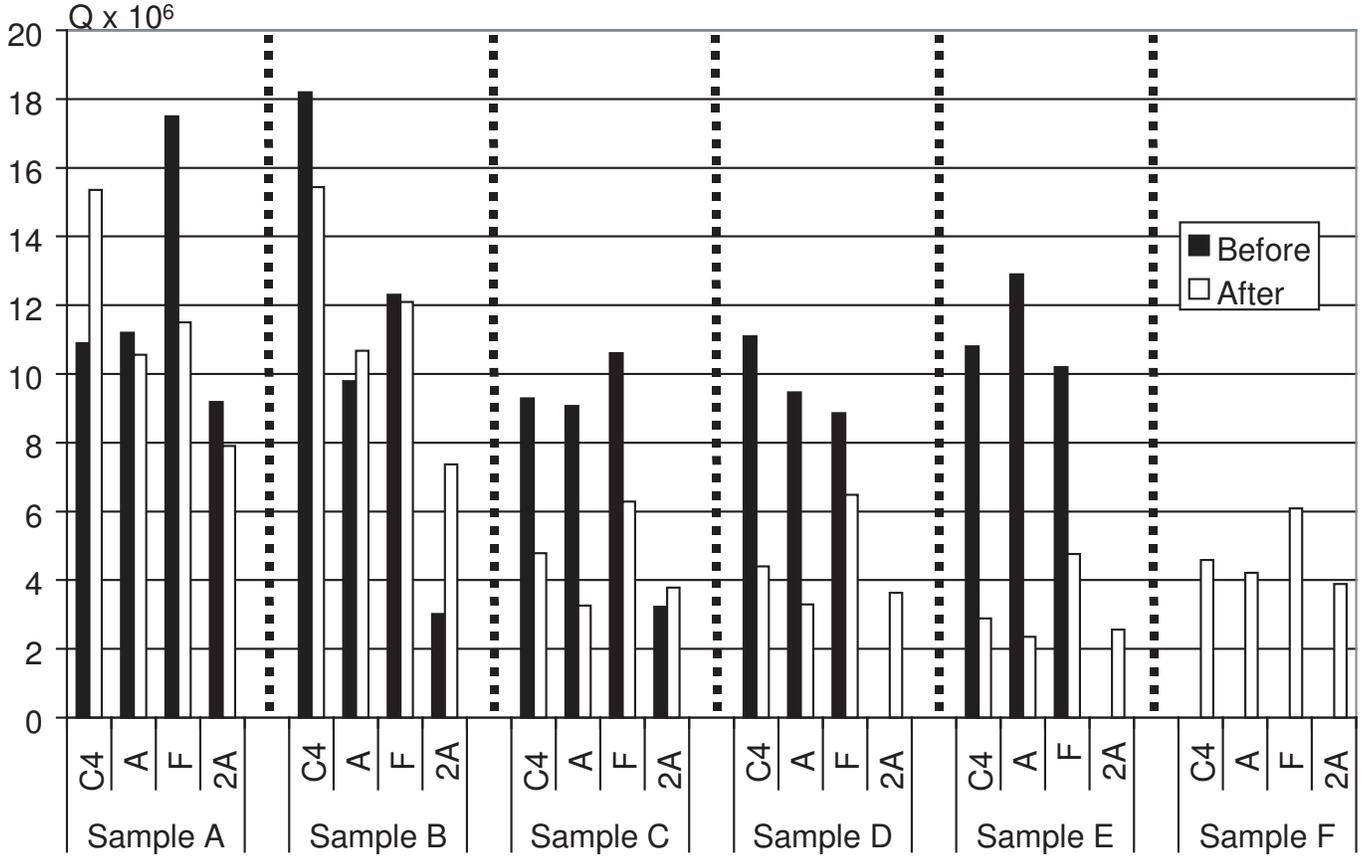} 
\caption{The measured $Q$ values for the four modes of each of the five thick samples with the coatings listed in Table~\ref{Tbl_SampleTypes}. The color of the bar indicates whether the data was measured before (black) or after (white) the coating/annealing process.}
\label{Fig_ThickResults}
\end{figure*}

\begin{figure}[htbp]
\includegraphics[width=8cm]{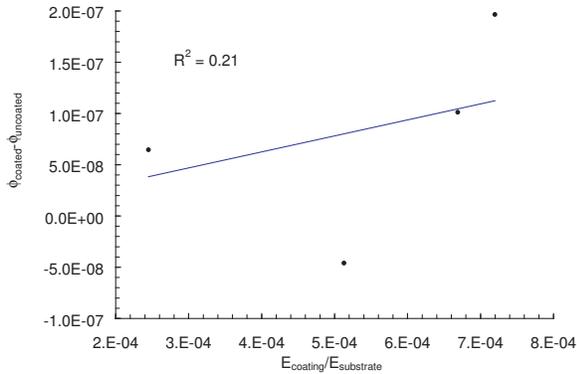} 
\caption{$\Delta\phi (f_0)$ plotted as a function of energy ratio for a typical sample.}
\label{Fig_CoatedInitial}
\end{figure}

\begin{figure}[htbp]
\includegraphics[width=8cm]{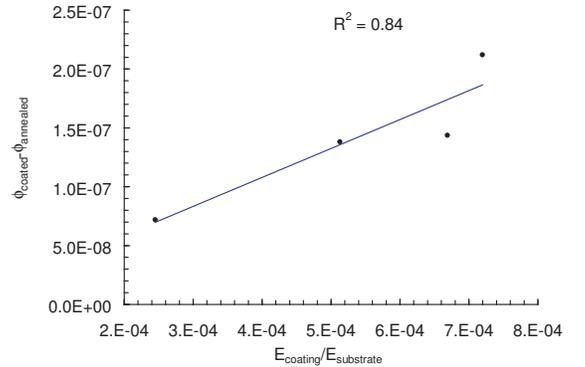} 
\caption{Same sample as in figure~\ref{Fig_CoatedInitial} except that $\Delta\phi (f_0)$ is corrected to account for the change
in substrate loss due to annealing.}
\label{Fig_CoatedAnnealed}
\end{figure}

Reviewing our results, listed in Table~\ref{Tbl_ThinResults}, we can test our
hypotheses on the source of the coating loss. The first hypothesis was that the
coating loss originated in the coating-substrate interface, in which case the
loss would be independent of coating thickness.  Were this idea true, then
coating types B and C, which have 2 and 30 layers respectively, would have
similar values for $Q_\mathrm{coated}$.  Instead, $Q_\mathrm{coated}$ for these
samples differ by a factor 10, and the coating loss, $\phi_\mathrm{coating}$,
which is scaled for coating thickness, is roughly equal for the two coating
types. Thus the loss is in the coating, not the coating-substrate interface.

Our second idea was that the coating loss arose in the interfaces between the
multiple coating layers.  To test this hypothesis we can compare the loss for
coating types C and D, which have the same total thickness but 30 and 60 layers
respectively.  If the loss were dependent on the number of coating interfaces
then $\phi_\mathrm{coating}$ for coating types C and D would differ by a
factor 2. However, coating types C and D have similar values for
$\phi_\mathrm{coating}$, indicating that the loss does not depend on the number
of coating interfaces.

Finally if the loss originates in the bulk coating material, then we would
expect to see a dependence in $\phi_\mathrm{coating}$ as the proportions of
coating materials are varied in coating types E, C, and F.  Indeed
$\phi_\mathrm{coating}$ does increase with increasing proportions of \Tan,
indicating that the loss is due to the bulk coating materials.

To separate $\phi_\mathrm{coating}$ into the losses for the two coating
materials, we must partition the energy in the coating into the amounts stored
in its silica and tantala layers.  Following an analysis by Landau and
Lifshitz~\cite{LaundauLifshitz}, it can be shown that, for Poisson's ratio,
$\sigma << 1$, the energy stored in a coating, $E_\mathrm{coating} \propto
{Ys}$, where $Y$ is Young's modulus and $s$ is the coating thickness. For a
multi-layer coating made up of two materials, we may write:
\begin{equation}\label{Eqn_Phi_12}
\phi_\mathrm{coating} = \frac{Y_1 s_1}{Y_\mathrm{coating} s_\mathrm{coating}} \phi_1 + \frac{Y_2 s_2}{Y_\mathrm{coating} s_\mathrm{coating}} \phi_2
\end{equation}
where $Y_i$, $s_i$ and $\phi_i$ are the Young's modulus, total thickness and
loss angle of the $i^\mathrm{th}$ coating material.  We use the Young's moduli
of $7.2 \times 10^{10}$ Pa for fused silica and $1.4 \times 10^{11}$ Pa for tantalum pentoxide~\cite{Martin}. 

For a thin surface layer, the stress will be predominantly parallel to the surface. In
this limit, the total Young's modulus for the multi-layer coating is given by 
\begin{equation}\label{Eqn_PhiTotal}
s_\mathrm{coating} Y_\mathrm{coating} = s_1 Y_1 +  s_2 Y_2
\end{equation}

Using eqn.~\ref{Eqn_PhiTotal}, we rewrite eqn.~\ref{Eqn_Phi_12} as
\begin{equation}\label{Eqn_Phi_r12}
\phi_\mathrm{coating} = \frac{s_1 Y_1 \phi_1 + s_2 Y_2 \phi_2}{s_1 Y_1 +  s_2 Y_2}
\end{equation}

Fitting the results for $\phi_\mathrm{coating}$ from coating types C, E, and F, we obtain the bulk loss values for silica and tantala
\begin{eqnarray*}
\phi_\mathrm{silica} = (0.1 \pm 0.1) \times 10^{-4} \mathrm{~~~(Butterfly~mode)} \\
\phi_\mathrm{tantala} = (4.7 \pm 0.1) \times 10^{-4} \mathrm{~~~(Butterfly~mode)}\\
\\
\phi_\mathrm{silica} = (-0.1 \pm 0.3) \times 10^{-4} \mathrm{~~~(Drumhead~mode)} \\ 
\phi_\mathrm{tantala} = (6.1 \pm 0.3) \times 10^{-4} \mathrm{~~~(Drumhead~mode)} \\
\end{eqnarray*}

As a final consistency check, we measured two samples with coating type C that
were coated by MLD~\cite{MLD}.  These measurements yielded
$\phi_\mathrm{coating} = 3.0 \times 10^{-4}$ for the butterfly mode, and
$\phi_\mathrm{coating} = 3.2 \times 10^{-4}$ for the drumhead  and
2$^{\mathrm{nd}}$ butterfly modes.  These results are in basic agreement with
the samples coated at SMA/Virgo~\cite{SMA} and indicate that the coating loss is
associated with the materials and not the coating process or manufacturer.

\subsection{Thick Sample Results}
\subsubsection{Initial observations}

For the thick fused silica samples (7.6 cm $\O \times 2.5$ cm thick), we
measured the  $Q$ factors of 4 modes both before and after each sample was
coated or annealed. The results of these measurements are shown in
figure~\ref{Fig_ThickResults}. The mode shapes, which were calculated using a
finite element package~\cite{ALGOR}, are shown in figure~\ref{Fig_ThickModes}.

We suspended the thick samples in a loop of silk thread. Each sample was
suspended multiple times, with the suspension length varied each time, and the
highest $Q$ factor for each mode was used in our analysis. Previous
experiments~\cite{Braginsky, Logan, Rowan} have shown that measured $Q$ factors
may be too low if the resonant frequency of a test mass happens to coincide
with a resonant frequency of the suspension wires. Using this technique we have
shown that suspension losses can be reduced to a negligible level when
measuring $Q$'s of the order of a few times $10^7$~\cite{Crooks}.

Before discussing the quantitative analysis of the results, it is instructive
to observe the trends in the data shown in figure~\ref{Fig_ThickResults}.
Firstly, it can be seen that the $Q$ factors for sample A, which was annealed,
and for sample B, which had a 2-layer coating, were very similar, and that
these $Q$ factors were considerably higher than the other samples which had
thicker coatings.  This difference indicates that the coating-substrate
interface is not the dominant source of mechanical loss.

Secondly, samples C and D, which have the same total coating thickness, but 30
and 60 coating layers respectively, have very similar $Q$ factors.  This result
suggests that the individual multi-layer interfaces are not the dominant source
of mechanical loss. 

Finally, the coatings on samples C, E, and F each have 30 layers and the same
total thickness, but vary the proportions of \Sil\ and \Tan. The measured $Q$'s
show that the loss increases with increasing proportions of \Tan\ in the
coating. These results suggest that the bulk coating materials are the dominant
source of mechanical loss in the coatings, and that tantalum pentoxide has a
higher mechanical loss than silica.

These observations are all consistent with the trends seen in the thinner
samples discussed in section~\ref{Sect_ThinResults}.

It is important to note that the measured $Q$ factors for the samples are
mode-dependent. This dependence will be discussed in the following section.

\subsubsection{Quantitative analysis}

Calculating the coating loss for the thicker samples requires a slightly
different method than for the thin samples. For the thin samples, the loss in
the substrate was negligible, so the coating loss could be calculated directly
from the measured loss in the coated sample. For the thick samples, a much
smaller fraction of the energy was stored in the coating. Thus the loss in the
substrate must be considered when calculating the loss in the coating. 

For each mode of a thick coated sample, we calculate the energy ratio
${E_\mathrm{coating}}/{E_\mathrm{substrate}}$.  We then rewrite
eqn.~\ref{Eqn_PhiCoated} as:

\begin{equation}\label{Eqn_DeltaPhi}
\Delta\phi = \phi_\mathrm{coated} - \phi_\mathrm{substrate} \approx \frac{E_\mathrm{coating}}{E_\mathrm{substrate}} \phi_\mathrm{coating}
\end{equation}

To obtain $\phi_\mathrm{coating}$, we plot $\Delta\phi$ versus
${E_\mathrm{coating}}/{E_\mathrm{substrate}}$ for each mode.
Figure~\ref{Fig_CoatedInitial} shows this plot for one sample with thirty
alternating layers of \Sil\ and \Tan. It can be seen that the data appears far
from the expected straight line.

From Figure~\ref{Fig_ThickResults} it can be seen that there were significant
changes in the $Q$ factors of modes of a sample which was annealed but not
coated. The $Q$'s of some modes increased while others decreased.  In our case,
the mechanism for changes in $Q$ is not known but we postulate it may be due to
some redistribution of stress in the samples, which may produce mode-dependent
losses~\cite{Numata}. Thus while using the measured values for the $Q$ of the
uncoated substrate seems invalid, replacing these values by the $Q$ factors of
the modes of the annealed, but uncoated mass, should in principle now allow us
to obtain a value for $\phi_\mathrm{coating}(f_0)$. The results of this are
shown in figure~\ref{Fig_CoatedAnnealed}.

It can be seen that there is a significant improvement of the fit to the data
using this analysis. Carrying out the same types of analyses on the data from
the other coated samples gave very similar results.

\subsubsection{Results}

The loss factors obtained for each type of coating studied are shown in
table~\ref{Tbl_ThickResults}.

\begin{table}[htb]
\begin{tabular}{|c|c|c|}\hline
\parbox{0.75 in}{\vspace{0.1 cm}\bf{Coating Type}\vspace{0.1 cm}} & \parbox{0.75 in}{\bf{Number of Samples}}   &  \parbox
{1.3 in}{\( \mbox{\boldmath $\phi_\mathrm{coating}$}\)} \\ \hline\hline
B & 1 & $(0.9 \pm 2.8) \times 10^{-4}$ \\ \hline
C & 2 & $(2.7 \pm 0.7) \times 10^{-4}$ \\ \hline 
D & 2 & $(2.7 \pm 0.5) \times 10^{-4}$ \\ \hline
E & 2 & $(3.7 \pm 0.5) \times 10^{-4}$ \\ \hline 
F & 2 & $(1.9 \pm 0.2) \times 10^{-4}$ \\ \hline 
\end{tabular}
\caption{Mechanical loss factors for the various coating types calculated from the thick sample data.}
\label{Tbl_ThickResults}
\end{table}

The results shown in table~\ref{Tbl_ThickResults} confirm our earlier
observations indicating that the dominant source of mechanical loss is
associated with the bulk coating materials, and also confirm that the tantalum
pentoxide component of the coatings has a higher mechanical loss factor than
the silica component. This last deduction can be treated more quantitatively.

As with the thin samples, we use eqn.~\ref{Eqn_Phi_r12} and the results for
$\phi_\mathrm{coating}$ from coating types C, E, and F, to calculate the loss
for the silica and tantala layers.  The results are:
\begin{eqnarray*}
\phi_\mathrm{silica} =  (0.5 \pm 0.3) \times 10^{-4} \\
\phi_\mathrm{tantala} =  (4.4 \pm 0.2) \times 10^{-4}
\end{eqnarray*}

We also performed a consistancy check on the thick samples by measuring two samples that had coatings of type C applied by MLD~\cite{MLD}.  The measured coating losses were
$\phi_\mathrm{coating} = 2.8 \times 10^{-4}$ for sample 1, and
$\phi_\mathrm{coating} = 3.8 \times 10^{-4}$ for sample 2.  Like the thin samples, the thick samples yielded similar results for the two coating manufacturers. In both cases the coatings from MLD~\cite{MLD} showed slightly higher loss than the equivalent coatings produced by SMA/Virgo~\cite{SMA}.  

\section{Conclusions}
For advanced interferometric gravitational wave observatories to be able to
reach realistic astronomical distances, low mechanical loss mirror coatings are
necessary.  Observatories currently in operation use a multi-layer dielectric
coating of silica and tantala.  We have measured the loss in this coating to be
$\phi = 2.7 \times 10^{-4}$, which is in basic agreement with the  results from our previous
work~\cite{Harry02,Crooks}. Those references discuss how this coating loss may effect the 
sensitivity of Advanced LIGO.
 We have shown in this work that the loss in this coating is
predominantly the bulk loss in the tantala layers.  
Work is underway to investigate other coating materials and processes that may
give lower mechanical loss while retaining the high reflectivity, low scatter,
and low optical loss required by these advanced detectors.
\\
\section{Addendum}

The coating applied to the thick disk as reported in reference~\cite{Harry02} is
incorrect.  The correct coating thickness of 4.66~$\mu$m results in a calculated
coating $\phi$ of $5.2 \times 10^{-4}$.

\end{document}